\documentclass[journal,transvt]{IEEEtran}
\usepackage{amsfonts}
\usepackage{dsfont}
\usepackage{floatrow} 
\usepackage{setspace}
\usepackage{color}
\usepackage[table]{xcolor}
\usepackage{amssymb}
\usepackage{cite}
\usepackage[cmex10]{amsmath}
\usepackage{algorithm, algpseudocode}
\usepackage{algorithmicx}
\usepackage{array} 
\usepackage{mathrsfs}
\usepackage{graphicx}
\usepackage{latexsym}
\usepackage{amscd}
\usepackage{amsfonts}
\usepackage{caption}
\usepackage{subcaption}
\usepackage{amsmath,amscd,amssymb,verbatim}
\usepackage{graphics}
\usepackage{amsthm}
\usepackage[T1]{fontenc}
\usepackage[utf8]{inputenc}
\usepackage{authblk}
\usepackage[short]{optidef}
\usepackage{mathtools}
\usepackage{soul}
\usepackage{textcomp}
\usepackage{xcolor}
\usepackage{datetime}
\usepackage{microtype}

\usepackage{blindtext}
\IEEEoverridecommandlockouts

\newtheorem{theorem}{Lemma}
\newtheorem{corollary}{Corollary}
\newtheorem{assumption}{Assumption}

\newcommand{\eqsplit}[1]{\begin{equation}\begin{split} #1 \end{split}\end{equation}}

\begin{document}
%
\renewcommand{\thefootnote}{\fnsymbol{footnote}}
\title{Fairness-aware Age-of-Information Minimization in WPT-Assisted \textcolor{black}{Short-Packet Data Collection for mURLLC}  
\thanks{
Part of this work, i.e., the optimization problem in~\eqref{problem:single} with a single variable, was presented at 2022 IEEE International Conference on Communications~\cite{Zhu_ICC_2022}.   

 Y. Zhu, X. Yuan, and Y. Hu are with School of Electronic Information, Wuhan University, 430072 Wuhan, China and   Chair of Information Theory and Data Analytics, RWTH Aachen University, 52074 Aachen, Germany. (Email:$zhu|yuan|hu$@inda.rwth-aachen.de). 

 B. Ai is with School of Electronic and Information Engineering, Beijing Jiaotong University, (Email: $boai$@bjtu.edu.cn).

 B. Han is with the Division of Wireless Communications and Radio Positioning, RPTU Kaiserslautern-Landau, 67663 Kaiserslautern, Germany (Email: bin.han@rptu.de).

R. Wang and A. Schmeink is with the Chair of Information Theory and Data Analytics, RWTH Aachen University, 52074 Aachen, Germany (Email: $wang|schmeink$@inda.rwth-aachen.de). 
 } }
\author{
Yao Zhu, 
Xiaopeng Yuan, 
Yulin Hu, 
Bo Ai, 
Ruikang Wang,
Bin Han,
Anke~Schmeink
}
\maketitle

\begin{abstract}
\textcolor{black}{The technological landscape is rapidly evolving toward large-scale systems. Networks supporting massive connectivity through numerous Internet of Things (IoT) devices are at the forefront of this advancement. In this paper, we examine Wireless Power Transfer (WPT)-enabled networks, where a server requires to collect data from these IoT devices to compute a task with massive Ultra-Reliable and Low-Latency Communication (mURLLC) services.} We focus on information freshness, using Age-of-Information (AoI) as the key performance metric. Specifically, we aim to minimize the maximum AoI among IoT devices by optimizing the scheduling policy. Our analytical findings demonstrate the convexity of the problem, enabling efficient solutions. We introduce the concept of AoI-oriented cluster capacity and analyze the relationship between the number of supported devices and network AoI performance. Numerical simulations validate our proposed approach's effectiveness in enhancing AoI performance, highlighting its potential for guiding the design of future IoT systems requiring mURLLC services.

\end{abstract}
\vspace*{-0.12cm}

\section{Introduction}
\label{sec:intro}

\textcolor{black}{In the dynamic world of technology, the trend is shifting towards large-scale systems, heralding a new era of possibilities in advanced networks. 
This shift has been particularly influential in the realm of  Internet of Things (IoT) applications, such as 
smart cities~\cite{Mehmood_smartcity_2017}, healthcare monitoring~\cite{Dressler_monitoring_2015},
and the Internet-of-nano-Things~\cite{Akyildiz_IonanoT_2010}. 
In the typical scenarios of these applications, the server is required to collect data from a massive number of IoT devices to compute the timely and accurate decision results.
Therefore, it is crucial to support seamless communications from countless IoT devices with rapid and robust connectivity.  
To address these challenges, the future 6G wireless technologies identify a key services class, massive Ultra-Reliable Low-Latency Communications (mURLLC), aiming to support stringent quality-of-service (QoS) requirements, such as ultra-reliability (greater than 99.9999\%), extremely low end-to-end delays (less than $1$ ms) while enabling massive connectivity~\cite{She_URLLC_intro_2021,Dang_mURLLC_2020}. 
Therefore, the finite blocklength (FBL) codes are likely to be employed. 
Unlike the well-known assumption of infinite blocklength, with FBL codes, data transmissions are no longer arbitrarily reliable, even the transmission rate is lower than the Shannon capacity~\cite{Polyanskiy_2010}.} 

\textcolor{black}{Despite the progress in these advanced services, a major challenge faced by IoT devices particularly is their limited energy storage due to their low-cost and simple circuits. This restriction poses significant
challenges for their sustained operation. Consequently, Wireless Power Transfer (WPT) emerges as a potential solution to this energy challenge~\cite{Huang_WPT_intro_2019}.} By leveraging radio-frequency (RF) energy harvesting (EH), IoT devices can convert RF signals into transmit power. Unlike conventional energy harvesting technologies that rely on ambient sources like heat, pressure, or vibrations, WPT provides a consistent and adjustable power source. This ensures stable performance for IoT devices under varying operational conditions.

Furthermore, considering the essential role of IoT devices in delivering real-time environmental data, the timeliness of information becomes crucial. 
The data relayed by these devices must be current, accurately reflecting the environment's state. 
To quantify the timeliness of the transmitted information, the concept of Age of Information (AoI) was introduced~\cite{Sun_AoI_intro_2017}, serving as a novel metric in wireless communications. 
In the context of IoT networks, it directly impacts the performance of real-time operations enabled by these networks. 
Therefore, managing AoI is essential to ensuring that the sensed data is not only accurate but also timely. 
Motivated by this background, various studies have been carried out in EH-enabled communication scenarios. 
However, how to minimize the fairness-aware AoI in the WPT-enabled IoT networks still remains unexplored, especially with the consideration of mURLLC. 
Moreover, how to design the resource allocation policy for information updates efficiently is also another concern, since the optimization of such policy often faces the issue of scalability to fulfill the needs of the enormous number of devices in IoT networks~\cite{Nguyen_IoT_intro_2022}. 
Determining the maximum number of devices a network can support without compromising data freshness is yet to be resolved. 

In this paper, we tackle the challenges mentioned above, \textcolor{black}{focusing on a fairness-aware optimal resource policy, which minimizes the maximum of the long-term AoI among devices in WPT-enabled networks for mURLLC services.}  
We formulate the optimization problem and convert the high-complexity approach to solve the problem to a more efficient one without impacting on the optimality.  
Furthermore, we introduce the concept of AoI-oriented cluster capacity and discuss the relationship between the number of supported devices and the AoI performance in the network. The
main contributions of this paper are:

\begin{itemize}
    \item \textbf{Providing a fairness-aware AoI minimization approach} for WPT-enabled networks supporting mURLLC services via optimizing the update schedule, considering the impact of FBL codes on time-average AoI. 
    \item \textbf{Establishing an equivalent low-complexity scheduling policy} that maintains optimal AoI performance. This is achieved by transforming the problem into a more manageable form and simplifying the analysis. 
    \item \textbf{Characterizing the (quasi-)convexity of AoI and the error probability} with respect to the charging duration and update duration. Such analytical findings indicate the reformulated problem can be efficiently solved as a convex problem. 
    \item \textbf{Proposing the concept of cluster capacity and a scalable algorithm} to obtain the update scheduling by defining the saturation of the cluster, in order to fulfill the needs of massive connectivity and provide guidance for practical cluster designs.  
    \item  \textbf{Demonstrating the advantages of \textcolor{black}{the}  proposed approach} via numerical simulations. The impacts of different parameters on the AoI performance are also discussed.
\end{itemize}

The remainder of this paper is organized as follows. Related
works are briefly reviewed in Section~\ref{sec:related}. Section~\ref{sec:system_model} provides the considered system model. In Section~\ref{sec:problem}, the optimization problem is formulated and its solutions are provided. Section~\ref{sec:capacity} discusses the concept of cluster capacity.
Section~\ref{sec:simulations} presents the simulation results and Section~\ref{sec:conclusion} 
gives the conclusions. 

\section{Related Works}
\label{sec:related}

\emph{Managing AoI in WPT-enabled sensor networks:} 
There are several works targeting AoI management in the general WPT-enabled sensor networks. For example, the author in~\cite{Krikidis_AoI_WPT_2019} studies the AoI performance for a single sensor with one-shot WPT energy management, i.e., transmitting the update with all harvested energy. 
The multi-sensor scenarios are investigated in~\cite{Gu_AoI_WPT_multi_2020}, where a joint optimization problem resource allocation and user scheduling in the frequency domain are formulated.  
The authors in~\cite{Liu_AoI_WPT_UAV_2022} further investigate the AoI performance with  UAV-assisted networks, where the ground users rely on the harvested energy from the UAV's wireless power to upload the data. A reinforcement learning approach is proposed to address the dynamical optimization problem, aiming at average AoI minimization.  
However, the results obtained by most of these existing works are based on the assumption of infinite blocklength, where the transmissions are arbitrarily reliable. 

\emph{mURLLC with FBL codes:} 
To characterize the transmission performance with FBL codes, the authors in~\cite{Polyanskiy_2010} derive a closed-form expression for the achievable transmission rate. Following this characterization, a set of optimal system designs has
been provided for mURLLC services. 
In particular, the authors in~\cite{Pan_URLLC_NOMA_OMA_2020} investigate the joint power and blocklength allocation for both orthogonal multiple access (OMA) and non-orthogonal multiple access (NOMA) schemes to minimize the error probability for FBL transmissions. 
Moreover, the resource allocation scheme for cell-free MIMO systems is studied in~\cite{Pan_URLLC_CF_2023}, where the lower bound on the achievable transmission rate is derived with imperfect channel state information (CSI). 
On the other hand, the authors in~\cite{She_URLLC_secure_2022} focus on the perspective of security in mURLLC, with the potential presence of the eavesdropper. The optimal power control policy is investigated in different CSI scenarios. However, characterizing 
 the performance of mURLLC with conventional  metrics for data transmissions, e.g., delay or throughput, may not be comprehensive for IoT applications, for which the freshness of the data is more important. 

\textcolor{black}{\emph{The Characterization and Optimization of AoI with FBL codes:} 
Unlike the AoI performance with infinite blocklength (IBL) codes, the impact of transmission error plays an important role in the characterization of AoI while blocklength allocation is investigated for the optimization problem. For example, the AoI violation probability is  characterized in~\cite{Sung_FBL_aoi_2022} and the peak AoI is analyzed in~\cite{Cao_FBL_aoi_2023}. However, both of them apply the linear approximation of FBL error probability model~\cite{Yu_FBL_aoi_approx_2020}, which may not be accurate for mURLLC services. 
In the pioneering work~\cite{zhang_FBL_NANO_2021}, the FBL performance is characterized for THz mURLLC in nanonetworks. The effective capacity is maximized via the proposed resource allocation scheme while establishing the FBL system model in nanonetworks. 
That being said, the AoI optimization with FBL codes, which is fundamentally important to keep the information fresh in WPT-enabled sensor networks, is still missing.} 

\section{System Model and Problem Statement}
\label{sec:system_model}
\subsection{System Description}
\textcolor{black}{Consider a network is operated with clusters. In a given cluster, a server  collects information update from $I$ devices continuously to compute a mission-critical task}, where its set is denoted as $\mathcal{I}=\{1,\dots, I\}$.  
Due to the limited size and practical reasons, these devices are mounted with capacitor-based energy storage instead of batteries. 
Therefore, the update operation fully relies on the wireless power transfer (WPT) via radio frequency from the server. Assume the server is in a full-duplex mode, which transmits the radio signals while receiving the updates from devices, i.e., with the simultaneously wireless information and power transmission (SWIPT). 
However, due to the simple circuit, the devices are operated in a half-duplex mode. In other words, its behavior is mutually exclusive from each other, so that it can either harvest the energy or transmitting the update, i.e., with wireless information transfer (WIT). Since devices require WPT to be carried out continuously to harvest the energy, there is no downlink communications in the occupied channel, i.e., the device receives neither acknowledgement (ACK) nor non-acknowledgement (NACK) from the server. 

Without loss of generality, we consider that devices act as active sources, i.e. they always generate the fresh message in each update. Based on the well-known principle in freshness-oriented system design, devices follow the Last-Come-First-Serve (LCFS) queuing policy with a queue length of $1$, which is optimal for the buffer of message sources~\cite{Yates_AoI_intro_2021}. 
Therefore, outdated messages will be simply discarded as long as a more fresh message arrives. In other word, the message update model can be equivalently considered as a fresh message being generated in each device when it is scheduled to update. 

To avoid collisions and waste of radio resources, the updates follow a mutual schedule policy $\pi$ and is known by every device.  
In particular, for each $k$-th update, the $i$-th device begins to harvest the RF energy at the time-domain symbol instance $t$ over a charging duration of symbol length $m_{c,i,k}$\footnote{Since we focus on the FBL performance in this work, for the convenience of notation, any time duration $t_{\text{duration}}$ is normalized to the relaxed symbol length $m$ with symbol duration $T_{\text{symbol}}$, i.e., $t_{\text{duration}}=mT_{\text{symbol}}\in\mathbb{R}_{\geq 0}$. Note that the symbol length is an integer. Therefore, The integer value of $m$ can be directly obtained by compared the integer neighbors of the relaxed one. }  and receives the harvested energy \textcolor{black}{$E_{c,i,k}$} . Afterwards, the device transmits the update via the same channel over the update duration of symbol length $m_{r,i,k}$. We assume the update is encoded into a single packet with the packet size of $D$ bits. Therefore, the start time of the $k$-th update can be indicated by an index that:
\begin{equation}
    \label{eq:starM_index}
    a_{i,k}(t)=
    \begin{cases}
        1,&~ \text{$k$-th update starts at $t$},\\
        0,&~ \text{otherwise}.
    \end{cases}
\end{equation} 
Then, the exact start time of the $k$-th update of device $i$  can be obtained via $\sum^T_{t=1}a_{i,k}(t)t$, where $T\to\infty$ is the furthest time index.   
To make the schedule policy $\pi$ online, it requires up-to-date information shared among each update. This may introduce significant implementation complexity and signaling overhead, leading to unaffordable energy consumption for the network. 
Therefore, in this work, we are interested in the consistent offline-scheduling policy, where $a_{\pi,n,k}(t)=a_{\pi,n,k+1}(t+m_{c,i}+m_{r,i})$, which requires no information exchange between device once the synchronization is done. 
In other words, each update for the same device will be carried out periodically with the same charging duration $m_{c,i}$ and update duration $m_{r,i}$ by dropping the index $k$. Then, the total duration of each update round of the device $i$ is given by
\eqsplit{M_i=m_{c,i}+m_{r,i}=\sum^\infty_{t=1}a_{i,k+1}(t)t-\sum^\infty_{t=1}a_{i,k}(t)t,~\forall k.    
}

{\color{black}
\subsection{WPT and WIT Channel Model} 
Both the WPT and WIT of the cluster are carried out via the same channel. 
We assume the channels experience independent and identically distributed small-scale fading $\hat{h}_i$ 
, which is quasi-stastic. In the other words, $\hat{h}_i$ is constant within each update round, and may vary in the next. For the large-scale fading, it suffers from a path-loss of ${d_i^{-\eta}}$, where $\eta$ is the path-loss exponent. Therefore, we denote $\hat{z}_i$ as the channel gain of the device $i$, which can be written as
\begin{equation}
    \textcolor{black}{\hat{z}_{i}} =\hat{h}^2_i{d_i^{-\eta}}.  
\end{equation} 
Assuming the device $i$ with the capacitor-based
energy storage, its harvested energy within the charging duration is given by\footnote{Without loss of generality, we consider the linear EH model in this work. However, other non-linear EH models can also be adopted as long as the harvested energy is continuous and increasing in the charging duration~\cite{Jornet_non_linear_EH_2012,Hu_non_liear_EH_2020}.}: 
\begin{equation}
    E_{c,i}=\mu \hat{z}_{i} p_c\cdot m_{c,i},
\end{equation}

where $\mu$ is the EH efficiency and $p_c$ is the transmit power of the server. After harvesting $E_{c,i}$, it transmits its update with a single data packet of $D$ bits in update duration $m_{r,i}$ to the server. 
With the transmission signal $x_i$, the received signal $y_i$ is given by:
\begin{equation}
    y_i=\sqrt{\hat{z}_ip_{r,i}}x_i+n_i,
\end{equation} 
where $n_i$ is the noise and $p_{r,i}$ is the transmit power of the device $i$. 
At the server side, via self-interference cancellation technologies, transmission and reception are enabled to be operated simultaneously. Due to the imperfect cancellation, the residual interference is not negligible.
We denote by $h_I$ the power gain of the residual loop interference. 
Therefore, with the update duration $m_{r,i}$, the SNR of signal $y_i$ at the server is given by:
\begin{equation}
    \gamma_i=\frac{\hat{z}_{i}\frac{E_{c,i}}{m_{r,i}}}{h_Ip_c}=z_{i}\frac{m_{c,i}}{m_{r,i}},
\end{equation}
where $z_{i}=\frac{\mu p_c\hat{z}_i}{h_Ip_c+\sigma^2}$ is defined as the time-wrapped channel gain with noise power $\sigma^2$.}

\subsection{Characterization of the Time-Average AoI} 
\begin{figure}[t!]
    \centering
    \includegraphics[width=0.6\linewidth,trim = 0 0 0 0]{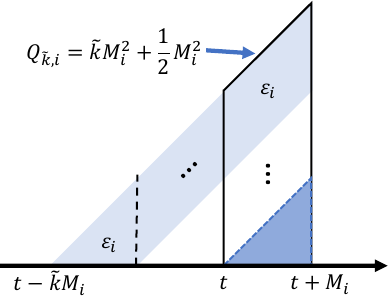}
    \caption{Evolution of AoI with the Event $X_{i,\tilde{k}}$.}
    \label{fig:aoi_evo}
\end{figure}

Since the purpose of the updates is to provide the current state of the environment, the conventional  metrics, e.g., delay or throughput, do not directly provide the freshness of the updates. Therefore, in this work, we consider a novel metric, age-of-information (AoI), to more \textcolor{black}{accurately}  characterize system performance~\cite{Sun_AoI_intro_2017}. In particular,  
the instantaneous AoI of the $i$-th  device represents the freshness of the update at symbol instance $t$, which is defined as:
\begin{equation}
    \Delta_i(t)=t-U_i(t),
\end{equation}    
where $U_i(t)$ is the instance of the most recent update that is \emph{successfully} received by the server from the device. 
Due to the impact of noisy channels and limited blocklength, it is determined by two factors: $1)$ how many failure update(s) before the current update round; $2)$ How long is each update round. 
Recall that the scheduling policy $\pi$ is consistent, and there is no feedback for the update. Then, the number of failure updates follows the Bernoulli process $\mathcal{B}(1-\varepsilon_i)$, where $\varepsilon_i$ is the packet error probability of each transmission. In particular, denote $X_{i,\tilde{k}}$ the event that the transmitted packet in the $\tilde{k}$-th update before the current update is decoded correctly while the next $\tilde{k}-1$ updates fail. Therefore, the probability of event $X_{i,\tilde{k}}$ can be written as:
\begin{equation}
    \mathsf{Pr}(X_{i,\tilde{k}})=\varepsilon_i^{\tilde{k}-1}(1-\varepsilon_i).
\end{equation}
Specially, we define $X_{i,0}$ as a sure event that the current transmission will occur with $\mathsf{Pr}(X_{i,0})=1$. 
Clearly, $\Delta_i(t)$ is a linear increasing function with respect to $t$ within $M_i$. Then, if the event $X_{i,\tilde{k}}$ occurs, the accumulated AoI contributed by such an event is represented by the area $Q_{i,\tilde{k}}$:
 \begin{equation}
    Q_{i,\tilde{k}}= \tilde{k}\textcolor{black}{M_i}^2+\frac{1}{2}\textcolor{black}{M_i}^2,
 \end{equation}
which are illustrated in Fig.~\ref{fig:aoi_evo}. 
Therefore, the expected time-average AoI $\bar{\Delta}_i$ is the sum of $Q_{i,\tilde{k}}$ from every possible event on the time infinite horizon:
\textcolor{black}{
\begin{equation}
\label{eq:aoi}
    \begin{split}
        \bar{\Delta}_i=\mathbb{E}[\Delta(t)]_i&=\lim_{t\to\infty}\frac{\sum^\infty_{\tilde{k}=1} \mathsf{Pr}(X_{i,\tilde{k}})Q_{i,\tilde{k}}}{t+M_i-t}\\&=\lim_{t\to\infty}\sum^\infty_{\tilde{k}=1}\left(\tilde{k}\textcolor{black}{+}\frac{1}{2}\right)\textcolor{black}{M_i}\left(\varepsilon^{\tilde{k}-1}_i-\varepsilon^{\tilde{k}}_i\right)\\
        &= \frac{1}{2}M_i+\sum^{\infty}_{{\tilde{k}}=1}{\tilde{k}}\varepsilon^{\tilde{k}}_iM_i\\
        &=\frac{1}{2}M_i+\frac{M_i}{1-\varepsilon_i},
        \end{split}
    \end{equation} 
}
The last equality holds as the error probability $\varepsilon_i\leq 1$. 

\subsection{Packet Error Probability in Finite Blocklength Regime}
Due to the limited update duration, the blocklength $m_{r,i}$ can no longer be considered as infinite. In other words, transmission error may still occur, even if the transmission rate is within the Shannon capacity region. In particular, with a given target error probability $\bar{\varepsilon}_i$, the maximal achievable rate in the FBL regime can be tightly approximated as~\cite{Polyanskiy_2010}:
\begin{equation}
    \label{eq:maximal_rate}
        r^*_i\approx \mathcal{C}(\gamma_i)-\sqrt{\frac{V(\gamma_i)}{m_{r,i}}}Q^{-1}(\bar{\varepsilon}_i),
    \end{equation}
where  ${\mathcal{C}}(\gamma_i) = {\log _2}( {1 + \gamma } )$ is the Shannon capacity and ~$V(\gamma_i)=1- {(1+\gamma_i)^{-2}}$ is the channel dispersion in AWGN channels~\cite{Chem_2015}. Moreover, $Q^{-1}(x)$ is the inverse Q-function with Q-function defined as $Q(x)=\int^\infty_x \frac{1}{\sqrt{2\pi}}e^{-\frac{t^2}{2}}dt$. Then, for any given packet size $d_i$ of the update, according to~\eqref{eq:maximal_rate}, the packet error probability of a single transmission can be written as:
\begin{equation}
\label{eq:error_tau}
    \textcolor{black}{\varepsilon_i \!\approx\! Q\Big( {\sqrt {\frac{m_{r,i}}{V(\gamma_{i})}} ( {{\mathcal{C}}(\gamma_i ) \!-\! \frac{d}{m_{r,i}})} \ln2} \Big)\mathrm{.}}
\end{equation}
Note that $\gamma_i$ depends on both charging duration $m_{c,i}$ and transmission duration $m_{r,i}$. Therefore, $\varepsilon_i$ is directly subject to the scheduling policy $\pi$.

\subsection{Problem Statement}
To maintain the fairness, in this work, we aim at minimizing the maximum of time-average AoI among devices by designing the policy $\pi$, including the update scheduling of the considered cluster, and update strategies of each device, i.e., $\pi=\{a_{i,k},m_{r,i},m_{c,i}~|~\forall t, i\}$. The corresponding optimization problem is as follows:
\begin{mini!}[2]
    {\pi}{\max_i~\{\bar{\Delta}_i\}}
    {\label{problem:original}}{}
    \addConstraint{\textcolor{black}{\sum_{j=1, j\neq i}^{I}}\sum^{m_{r,i}-1}_{\tau=0}a_{j,k}(t+\tau)\leq 1, \forall k \label{con:schedule}}
    \addConstraint{\varepsilon_{i}\leq \varepsilon_{\max},~\gamma_i\geq \gamma_{\text{th}}, \forall i\label{con:threshold_original}}
    \addConstraint{a_{i,k}(t)\in \{0,1\},~\forall,i,k,t\label{con:int},}
\end{mini!}
where the constraint~\eqref{con:schedule} avoids any transmission collision between the updates. The constraint~\eqref{con:threshold_original} ensures the quality of the updates that prevents the waste of resources, where $\varepsilon_{\max}\leq 0.5$ and $\gamma_{\text{th}}\geq 1$ the error probability and SNR threshold, respectively. 

With this optimization problem, in what follows, we intend to answer two key research questions:
\begin{enumerate}
    \item With any given device set $\mathcal{I}$, what is the optimal scheduling policy $\pi^*$ in the considered cluster with the consideration of fairness?
    \item Suppose there is already a group of devices, how many additional devices the cluster could support without influencing the freshness performance of the already existing devices? 
\end{enumerate}


\section{Problem Reformulation}
\label{sec:problem}
In this section, we reformulate the original problem into a more tractable one by converting the scheduling policy into a resource allocation policy. Then, we investigate the convexity of the reformulated problem by establishing the (quasi-)convexity of the FBL error probability and time-average AoI. Finally, after efficiently obtaining the optimal resource allocation policy, we reconstruct it back to the scheduling policy.  
\subsection{Problem Reformulation}
Clearly, Problem~\eqref{problem:original} is an integer non-convex problem. Although it can be solved via exhaustive search by upper-bounding the time horizon, i.e., $t\leq T_{\max}$, it is practically impossible to do so in large-scale IoT networks, since the complexity scales exponentially with the number of supported devices in the considered cluster. 

To this end, we reformulate the problem into an equivalent one, yet with a time- and order-independent resource allocation policy, that with significantly lower complexity. 
In particular, we first establish the following lemma:
\begin{theorem}
    \label{lemma:start_time}
    With a fixed update strategy $M_i=m_{c,i}+m_{r,i}$, the start time of the update $a_{i,k}(t)$ does not influence the time-average AoI $\bar{\Delta}_i$ over the time infinite horizon, i.e., $t\to\infty$.
\end{theorem}
\begin{proof}
    Let $t=m_{c,i}$ to be the first update round of a given policy $\pi$. Suppose that there is another policy $\pi'$ shifting the transmission start time by a duration of $t'\in[0, M_i)$ so that $a'_{i,k}(t-t')=a_{i,k}(t)$. With the consistent charging and transmission duration, it still holds that $a'_{i,k}(t)=a'_{i,k+1}(t)$.
    Then, the corresponding time-average AoI is given by: 
    \begin{equation}
        \begin{split}
      \bar{\Delta}'_i&=\lim_{t\to\infty}\bigg(\frac{\sum^\infty_{\tilde{k}=1} \mathsf{Pr}(X_{i,\tilde{k}})Q_{i,\tilde{k}} }{t+M_i-t}+\mathsf{Pr}(X_{i,\tilde{\infty}})\frac{\sum^t_{\tau=1} a_{i,1}(\tau)\tau}{t}\bigg)\\
      &=\lim_{t\to\infty}\sum^\infty_{\tilde{k}=1}\left(\tilde{k}-\frac{1}{2}\right)T\left((\varepsilon')^{\tilde{k}-1}-(\varepsilon')^{\tilde{k}}\right)
        \\
        &=\frac{1}{2}M_i+\frac{M_i}{1-\varepsilon'_i},
        \end{split}
    \end{equation}
    where $X_{i,0}$ is the event that $\varepsilon'_i$ is the packet error probability with the new policy. Note that even if the transmission start time is shifted, the overall charging duration $m'_{c,i}$ is still the same with consistent transmission duration $m_{r,i}$, i.e., 
    \begin{equation}
        m'_{c,i}=\sum^\infty_{t=1}a'_{i,k}(t)t-\sum^\infty_{t=1}a_{i,k}(t)(t-M_i+m_{r,i})=m_{c,i}.
    \end{equation}
    Straightforwardly, we can deduce $\bar{\Delta}'_i=\bar{\Delta}_i$ with $\varepsilon'_i(m'_{c,i},m_{r,i})=\varepsilon_i(m_{c,i},m_{r,i})$. 
    \end{proof}
Lemma~\ref{lemma:start_time} implies that we can arbitrarily choose when to start the update transmission within each update interval $[t, t+M_i)$ while keeping the same AoI with a consistent scheduling policy, $a_{n,k}(t)=a_{n,k+1}(t+M_i)$.   
Then, we make the following assumption:
\begin{assumption}
\label{assumption:M}
There is a consistent update duration $M$, which is feasible in Problem~\eqref{problem:original}, so that, within any interval $[\tau, \tau+M)$, each device updates once and once only, i.e., $\sum^{\tau+M-1}_{t=\tau}a_{i,k}(t)=1$.   
\end{assumption}
\emph{\textbf{Remark 1:} This may seem to be a strong assumption at first glance, since it forces the update round of every device in the cluster to be unified. However, surprisingly, this assumption does not influence the optimal solutions of Problem~\eqref{problem:original}, the proof of which will be shown in a later section.}

\begin{figure}[t!]
    \centering
    \includegraphics[width=0.9\linewidth,trim = 0 0 0 0]{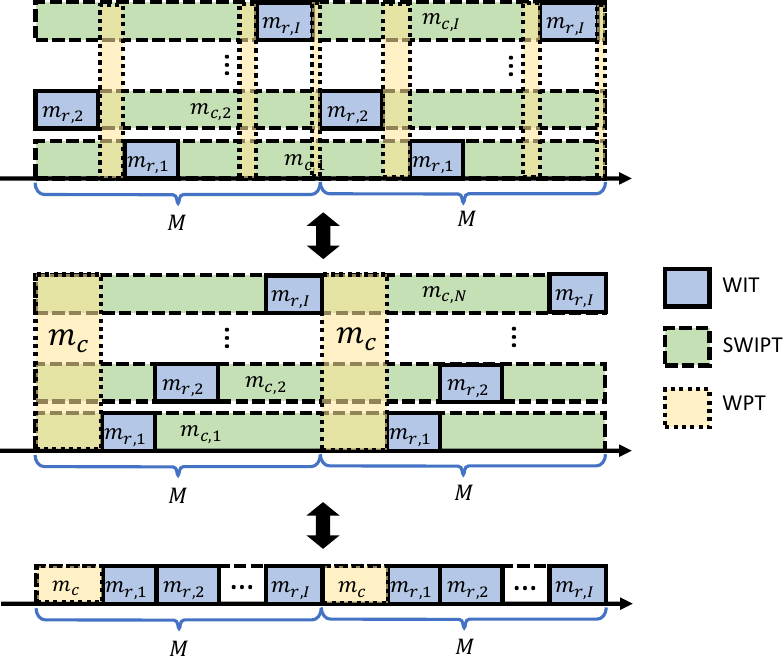}
    \caption{Equivalent update scheduling policies.}
    \label{fig:aoi_frame}
\end{figure}
Assumption~\ref{assumption:M} indicates that we can re-organize the $k$-th updates of each device together within a unified total duration $M$.  
Recall that there is no overlapping between update transmissions due to the constraint~\eqref{con:schedule}. Therefore, it may also exist certain instances at which no one transmits its update, i.e., every device is harvesting the energy. We define the sum of those instances as the common charging duration $m_c$. 
Then, according to Lemma~\ref{lemma:start_time}, any scheduling policy $\pi$ with a given $M$ that following Assumption~\ref{assumption:M}, can be equivalent to a time- and order-independent resource allocation policy, $\hat{\pi}=\{m_{r,i},m_{c,i}~|~m_{r,i}+m_{c,i}= M,~\forall i\in\mathcal{I}\}$. Moreover, it must hold that:
\begin{equation}
    m_{c,i}=\sum^I_{j \neq i} m_{r,j}+m_c. 
\end{equation}
In other words, $\hat{\pi}$ can be graphically interpreted as follows: At the beginning of every update round with a duration of $M$, a WPT phase is carried out with a duration of $m_c$ for every device. Then, the WIT of each device is carried out one by one while the rest of them keep harvesting the energy in the SWIPT phase.  
Without loss of generality, we consider the order of transmission follows the order of the device index. It should be emphasized that the actual transmission order does not matter, since the exchange of the index of any two devices has no impact on the average AoI performance, as we showed in~\cite{Han_AoI_2021}. The equivalence relation of the scheduling policy and resource allocation policy is shown in Fig.~\ref{fig:aoi_frame}.
Moreover, to replace the maximum in the objective function, we introduce a new variable $\Delta_{\max}$, which holds that $\Delta_i\leq \Delta_{\max},~\forall i$. 
Therefore, 
Problem~\eqref{problem:original} can be reformulated as: 
\begin{mini!}[2]
    {_{m_c,m_{r,1},\dots,m_{r,I},\Delta_{\max}}}{\Delta_{\max}}
    {\label{problem:reformulation}}{}
    \addConstraint{\bar{\Delta}_i\leq \Delta_{\max},\forall i\in\mathcal{I}\label{con:delta_max}}
    \addConstraint{m_{c,i}+m_{r,i}=M,\forall i\in\mathcal{I}\label{con:M}}
    \addConstraint{ m_{c,i}=\sum^I_{j \neq i} m_{r,j}+m_c, \forall i\in\mathcal{I}\label{con:charging}}
    \addConstraint{\varepsilon_{i}\leq \varepsilon_{\max},~\gamma_i\geq \gamma_{\text{th}}, \forall i\in\mathcal{I},\label{con:threshold}}
\end{mini!}
where we transfer the original objective function in~\eqref{problem:original} into a new objective function $\Delta_{\max}$ and $I$ constraints in~\eqref{con:delta_max}. Moreover, constraint~\eqref{con:M} ensures that the update round of every device is unified. Constraint~\eqref{con:charging} indicates that other devices are able to harvest the energy while one device is transmitting the update. 

Such a reformulation implies that to optimize the scheduling policy $\pi$ is to optimize the allocated updated duration of each device $m_{r,i}$, as well as the common harvesting duration $m_c$, i.e., $\pi=\{m_c,m_{r,1},\dots,m_{r,I}\}$. Unfortunately, Problem~\eqref{problem:reformulation} is still non-convex. To this end, we investigate the optimization framework to efficiently solve it. 

\subsection{Optimal Solutions of Problem~\eqref{problem:reformulation}}




In order to solve Problem~\eqref{problem:reformulation}, we first establish the following lemma:
\begin{theorem}
    \label{lemma:err_convex}
    $\varepsilon_i$ is convex in $m_{c,i}$ and $m_{r,i}$ within the feasible set of Problem~\eqref{problem:reformulation}, if 
    \eqsplit{
        \mathcal{C}m_{r,i}+3 d\geq \frac{4}{\ln(2)}~\text{and}~r_i\geq \frac{16-18ln(1+\gamma_i)}{87-12ln2}. \label{eq:convex_condition}
    }
\end{theorem}
\begin{proof}
    In Appendix~\ref{app:Lemma_convex}.
\end{proof}

\emph{\textbf{Remark 2:} Although the convexity feature characterized in Lemma~\ref{lemma:err_convex} depends on the condition~\eqref{eq:convex_condition}, it can be fulfilled in the region of interest of most practical applications. For example, it is fulfilled if the packet size $d\geq 3$ bits. In the remainder of the paper, we assume that the condition
~\eqref{eq:convex_condition} is implicitly fulfilled.}

\emph{\textbf{Remark 3:} Compared with the similar results in the existing works, e.g.,~\cite[Proposition 2]{Zhu_swipt_2019},~\cite[Proposition 3]{Cao_AOI_FBL_2021}, the results Lemma~\ref{lemma:err_convex} is stronger. First, it characterizes the joint convexity instead of partial convexity with respect to $m_{c,i}$ and $m_{r,i}$. Moreover, the condition in~\eqref{eq:convex_condition} is tighter and more practical, compared to other conditions.} 

Note that $m_{c,i}$ is the linear combination of $m_c$ and all $m_{r,j},~\forall j\neq i$. 
Lemma~\ref{lemma:err_convex} indicates that $\varepsilon_i$ is jointly convex in all optimization variables in Problem~\eqref{problem:reformulation}. This result helps us to characterize the convexity of the problem by establishing the following corollary:
\begin{corollary}
    \textcolor{black}{Under the same condition as in Lemma 2,} Problem~\eqref{problem:reformulation} is convex.
    \label{corollary:problem_convex}
\end{corollary}
\begin{proof}
    First, the objective function is affine i.e., convex.
    Then, we investigate the convexity of constraint~\eqref{con:delta_max}, where $\bar{\Delta}_i$ is involved. 
    In particular, we can reformulate $\bar{\Delta}_i$ as follows:
    \eqsplit{
    \bar{\Delta}_i
    =\frac{1}{2}M_i+\frac{M_i}{1-\varepsilon_i}
    =\frac{(m_{c,i}+m_{r,i})}{\frac{2(1-\varepsilon_i)}{3-\varepsilon_i}}
    \triangleq \frac{p(\boldsymbol{m})}{q(\varepsilon_i(\boldsymbol{m}))},
    }        
    where $\boldsymbol{m}=\{m_c,m_{r,1},\dots,m_{r,I}\}$ is the variable vector of Problem~\eqref{problem:reformulation}. 
    Moreover, we can directly show that $q(\varepsilon_i)$ is concave and decreasing in $\varepsilon_i$ with:
    \eqsplit{
        \frac{\partial q}{\partial \varepsilon_i}=-\frac{4}{(\varepsilon_i-3)^2}\leq 0.
    }
    \eqsplit{
        \frac{\partial^2 q}{\partial \varepsilon_i^2}=\frac{8}{(\varepsilon_i-3)^3}\leq 0.
    }
    The above inequality holds since we have $0\leq \varepsilon_i\leq 1$. 
    According to Lemma~\ref{lemma:err_convex}, $\varepsilon_i$ is convex in $\boldsymbol{m}$. 
    Therefore, as a positive composition function, $q(\varepsilon_i(\boldsymbol{m}))$ is convex~\cite{boyd_convex_2004}. 
    Moreover, it is also trivial to show that $p(\boldsymbol{m})$ is a linear and convex function. Then, $\bar{\Delta}_i=\frac{p(\boldsymbol{m})}{q(\varepsilon_i(\boldsymbol{m}))}$, as a convex-over-concave function, is quasi-convex, i.e., constraint~\eqref{con:delta_max} is convex. 
    The rest of the inequality constraints are either affine or convex while all equality constraints are affine. 
    Hence, Problem~\eqref{problem:reformulation} is convex.
\end{proof}

Based on Corollary~\ref{corollary:problem_convex}, Problem~\eqref{problem:reformulation} can be solved efficiently with any standard convex optimization tools with a computational complexity of $\mathcal{O}((I+1)^2)$. Then, with its optimal solutions $(m^*_c,m^*_{r,1},\dots,m^*_{r,I})$, we can set the start time of $k$-th update of device $i$ as 
\begin{equation}
    \label{eq:a_k_optimal}
    a^*_{i,k}(t)=
    \begin{cases}
        1, & t=(k-1)M+m^*_c+\sum^i_{j=1}m^*_{r,i-1},\\
        0, & \text{otherwise},
    \end{cases}
\end{equation}
with which we can reconstruct the corresponding scheduling policy $\pi^*$ directly. 

Therefore, we are able to answer the first key research question in Section~\ref{sec:system_model}: The optimal scheduling policy $\pi^*$ can be obtained by solving a reformulated convex Problem~\eqref{problem:reformulation} with an equivalent resource allocation policy as shown in Fig~\ref{fig:aoi_frame}.   

\section{Cluster Capacity and Efficient Solutions}
\label{sec:capacity}
Although convex programming is well-known for its efficiency, for massive connectivity, we still face the scalability issue, since the computational complexity increases in the number of devices. Therefore, we are interested in a more efficient approach to obtain the scheduling policy. Moreover, it does not provide any technical insights for the system design of the cluster by solely solving Problem~\eqref{problem:reformulation} as a convex one. To this end, in this section, we further investigate our system from another perspective by introducing the concept of cluster capacity. Based on that, we also propose a low-complexity approach to obtain the scheduling policy. Finally, we analytically confirm that Assumption~\ref{assumption:M} is valid for the optimal scheduling policy.

\subsection{Cluster Capacity}
As discussed in the previous section, with the given device set $\mathcal{I}$, we are able to obtain the optimal scheduling policy $\pi^*$ including an optimal and common charging duration $m^*_c$. Interestingly, if $m^*_c$ is non-zero, we have the following observation: 

\begin{corollary}
    \label{corollary:mono}
    With any $\min_{i} \{m^*_{r,i}\}\leq m^*_c$, $\forall i \in \mathcal{I}$, we can always introduce an additional device with the same or better channel gain of device $i_{\min}=\arg \min_i \{z_i\}$ into the cluster without influencing the minimized $\Delta^*_{\max}$. 
\end{corollary}
\begin{proof}
    Let $I+1$ be the index of the additional devices. Then, its channel gain must fulfill $z_{I+1}\geq z_{i^*_{\min}}$. It is clear that its average AoI $\bar{\Delta}_{I+1}$ is monotonically decreasing in $z_{I+1}$, since we have:
    \eqsplit{
        \frac{\partial \bar{\Delta}_{i}(z_{i}|m^*_{c,i},m^*_{r,i})}
            {\partial z_i}
        =\underbrace{\frac{\partial \bar{\Delta}_{i}}
        {\partial \gamma_i}}_{\leq 0}
        \underbrace{\frac{\partial \gamma_i}
        {\partial z_i}}_{\geq 0}\leq 0.
    }
    It means that, with a given charging and update duration $(m^*_{c,i},m^*_{r,i})$, it always holds $\bar{\Delta}_{i}(z_{I+1}|m^*_{c,i},m^*_{r,i})\leq \max_i{\bar{\Delta_{i}}}\leq \Delta^*_{\max}$. 
\end{proof}

Therefore, the common charging duration $m_c$ can be viewed as the remained "free space" of the corresponding scheduling policy $\pi$, within which other devices may transmit their update with no AoI performance cost of the cluster. 
Then, suppose that there is only one device $i$ with channel gain $z_i$ in the cluster, we can obtain the optimal scheduling policy by solving the following optimization problem:
\begin{mini!}[2]
    {_{m_{c,i},m_{r,i}}}{\bar{\Delta}_i}
    {\label{problem:single}}{}
    \addConstraint{m_{c}=m_{c,i}}
    \addConstraint{\varepsilon_{i}\leq \varepsilon_{\max},~\gamma_i\geq \gamma_{\text{th}}, \forall i\label{con:threshold}}
    \addConstraint{(m_{c},m_{r,i})\in\mathbb{R_+},~\forall i\label{con:relax},}
\end{mini!}
which is clearly convex according to Corollary~\ref{corollary:problem_convex}, and therefore it can be \textcolor{black}{solved} efficiently.  
Denote its optimal solution as $m^\circ_{c,i}+m^\circ_{r,i}=M^\circ$ and minimized AoI $\bar{\Delta}^\circ_i$, with which we can obtain the corresponding scheduling policy $\pi^\circ$. 
Similar to the observation in Corollary~\ref{corollary:mono}, under the optimal solutions, $m^\circ_{c,i}$ is the largest "free space" of the cluster with a given device $i$, within which other devices may transmit their update without influencing minimized $\bar{\Delta}^\circ_i$. As discussed in Section~\ref{sec:system_model}, we are interested in the total number of devices that can be supported in the cluster with the consideration of fairness. This can be addressed by quantifying the "free space" with the following definition in terms of the number of devices.  


\emph{\textbf{Definition 1:} 
The fairness-aware \underline{cluster capacity} of a given set $\mathcal{I}$ is the maximal number of devices can be introduced in the cluster so that the minimized maximum AoI is not higher than the minimized AoI in a single-device cluster with the worst channel gain $z_{i_{\min}}\leq z_i$, $\forall i\in\mathcal{I}$, i.e.,
\begin{equation}
   \Delta^*_{\max}(\pi^*|\mathcal{I})\leq  \bar{\Delta}^\circ_{i_{\min}}(m^\circ_{c,i_{\min}},m^\circ_{r,i_{\min}}|z_{i_{\min}}).
\end{equation}
It can be expressed as
\begin{equation}
    \label{eq:capa}
    C_{\text{cap}}(\mathcal{I})=\lfloor\frac{m^\circ_{c,i_{\min}}+m^\circ_{r,i_{\min}}}{m^\circ_{r,i_{\min}}}\rfloor.
\end{equation}
}

It should be emphasized that the cluster capacity with the consideration of fairness is relative to the reference channel gain of the given set instead of an absolute quantity. 
Interestingly, according to Corollary~\ref{corollary:mono}, $C_{\text{cap}}$ can be obtained by sorting the channel gains in $\mathcal{I}$ and solving Problem~\eqref{problem:single} once. This is due to the fact that the performance of the cluster is bounded by the worst channel gain $z_{i_{\min}}$. 
However, there is no guarantee that the cluster capacity is able to cover the needs of the set. In other words, it is possible that the number of devices in the set exceeds its capacity, i.e., $C_{\text{cap}}(\mathcal{I})< |\mathcal{I}|$.  
To tackle this issue, we establish the next definition. 

\emph{\textbf{Definition 2:} The cluster is considered as \underline{saturated} under a scheduling policy $\pi$, if there is no common charging duration, i.e., $m_c=0$.}


Similarly to the definition of cluster capacity, the saturation of the cluster is also relative to the scheduling policy $\pi$.  In fact, according to Corollary~\ref{corollary:mono}, for the optimal scheduling policy $\pi^*$ obtained by solving Problem~\eqref{problem:reformulation}, the cluster is always saturated if $C_{\text{cap}}(\mathcal{I})< |\mathcal{I}|$. 

Interestingly, with the help of these definitions, we are actually able to re-\textcolor{black}{examine}  Assumption~\ref{assumption:M} with the following lemma.

\begin{theorem}
    \label{lemma:confirm_assumption}
    Within any interval $[\tau, \tau+M^*)$, each device transmits its update once and once only, i.e., $\sum^{\tau+M^*-1}_{t=\tau}a_{i,k}(t)=1$, in the optimal scheduling policy $\pi^*$. 
\end{theorem}

\begin{figure}[t!]
    \centering
    \includegraphics[width=1\linewidth,trim = 0 0 0 0]{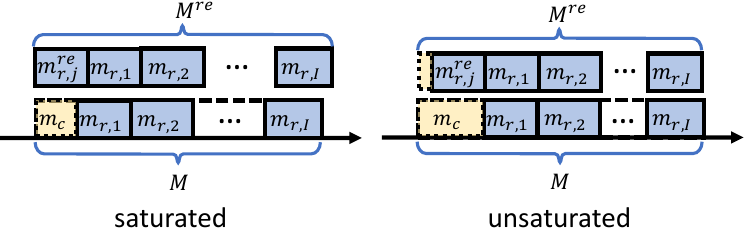}
    \caption{The impact of an additional update to the system in both saturated and unsaturated case with an additional update duration $m^{re}_{r,j}$.}
    \label{fig:saturated}
\end{figure}

\begin{proof}
    This can be proven by contradiction. 
    Suppose that there exists another scheduling policy $\pi^{re}$ with second update for any device with index $j$, that improves the minimized maximum AoI $\Delta^*_{\max}$ for an optimal scheduling policy $\pi^*$ with no second update, i.e., $\Delta^{re}_{\max}<\Delta^*_{\max} $. Let $m^{re}_{r,j}$ be the second update duration. Then, with the new duration of each update round $M^{re}$, we have the following two cases as shown in Fig.~\ref{fig:saturated}:
    \begin{itemize}
        \item If $m_c < m^{re}_{r,j}$, the cluster is saturated under $\pi^{re}$ with any second update. It must hold that $M^{re}=\sum^I_{i=1} m^*_{r,i}+m^{re}_{r,j}>M^*$. Since $\pi^*$ is optimal with the single update, we have $\bar{\Delta}_{i_{\min}}(m^{re}_{c,i},m^*_{r,i})\geq \Delta_{i_{\min}}(m^*_{c,i},m^*_{r,i})$ according to Corollary~\ref{corollary:mono}, where ${i_{\min}}=\arg \min_i\{z_i\}$ is the index of device with the worst channel gain. Recall that $\bar{\Delta}_{m_{c,i},m_{r,i}}\leq \Delta_{\max}$. Then, we can conclude that $\Delta^{re}_{\max}\geq \Delta^*_{\max}$. Therefore, the assumption of the improvement with the second update is violated. 
        \item If $m_c \geq m^{re}_{r,j}$, the cluster is unsaturated under $\pi^{re}$ with any second update. Then, we have $M^{re}=M^*$ and $m^{re}_{c}=m^*_c-m^{re}_{r,j}$. Since $\pi^{re}$ improves $\Delta^*_{\max}$ and the cluster is unsaturated, $j$ must be $i_{\min}$ and $\bar{\Delta^{re}}_{i_{\min}}<\bar{\Delta^*}_{i_{\min}}$. However, if $j$ is $i_{\min}$, the cluster must be saturated, otherwise $\bar{\Delta}_{i_{\min}}$ cannot be improved according to Definition 1. It violates the assumption of the unsaturated cluster. 
    \end{itemize} 
As result, both cases violate the given assumption. 
\end{proof}
Lemma~\ref{lemma:confirm_assumption} confirms that Assumption~\ref{assumption:M} indeed matches the optimal scheduling policy. Therefore, the optimal solutions of the reformulated problem in~\eqref{problem:reformulation} are equivalent to the optimal solutions of the original Problem in~\eqref{problem:original}.  

\subsection{Low-complexity Algorithm for solving Problem~\eqref{problem:reformulation}}
With the concept of cluster capacity and saturation,  
we further propose an efficient approach to obtain the scheduling policy in Problem~\eqref{problem:reformulation} to improve the scalability performance in networks. 

In particular, with any given set $\mathcal{I}$, we first sort the channel gain to find the worst channel gain $z_{i_{\min}}$. Then, we solve Problem~\eqref{problem:single} to get the optimal solution $(m^\circ_{c,\min},m^\circ_{c,\min})$. 
If the number of devices does not exceed its capacity, i.e., $C_{cap}(\mathcal{I})\geq I$, for any other device $j$, where $j\neq i$ and $j\in\mathcal{I}$, we minimize its average AoI $\bar{\Delta}_i$ with an additional constraint on the duration of its update round $m_{r,j}+m_{c,j}=M^\circ$. The problem is given by:
\begin{mini!}[2]
    {_{m_{c,j}, m_{r,j}}}{\bar{\Delta}_j}
    {\label{problem:single_j}}{}
    \addConstraint{m_{r,j}+m_{c,j}=M^\circ\label{con:M_j}}
    \addConstraint{\varepsilon_{j}\leq \varepsilon_{\max},~\gamma_j\leq \gamma_{\text{th}}}
    \addConstraint{(m_{c,j},m_{r,j})\in\mathbb{R_+}.}
\end{mini!}
Clearly, Problem~\eqref{problem:single_j} is also convex and can be solved efficiently, since the additional constraint~\eqref{con:M_j} is affine. 
Denote the obtained solutions as $(m'_{c,j},m'_{r,j})$, which implies the optimal scheduling policy of a single device $j$ with a fixed and unified duration of the update round. 
It should be emphasized that this scheduling policy is not necessarily optimal if we relax the constraint~\eqref{con:M_j}. In other words, the optimal solutions of Problem~\eqref{problem:single} and Problem~\eqref{problem:single} may differ unless the device $j$ has the same channel gain as the one of the device $i_{\min}$, i.e., $z_j=z_{i_{\min}}$. 
After solving $I-1$ convex problems, we can construct the scheduling policy for the whole set $\mathcal{I}$ by letting $m^\circ_{r,j}=\max\{m'_{r,j}, m^\circ_{r,i}\}$,~$\forall j\in\mathcal{I}$, which ensures the fairness performance. Then, 
the charging duration of each device is $m^\circ_{c,i}=M^\circ-m^\circ_{c,i}$. Moreover, the start time of each update $a^\circ_{i,k}$ can be obtained with~\eqref{eq:a_k_optimal}. 
Therefore, the scheduling policy is $\pi^\circ=\{a^\circ_{i,k}, m^\circ_{c,i},m^\circ_{r,i},\}$. According to Corollary~\ref{corollary:mono}, it achieves the globally optimal solutions in Problem~\eqref{problem:reformulation}, but only with a low computational complexity of $\mathcal{O}(I\log I+4I)$, since it requires a sorting for $I$ elements and solving $I$ independent convex optimization problems with two variables. 

However, if the number of devices exceeds its capacity, i.e., the cluster is saturated with $C_{cap}(\mathcal{I})\geq I$, we can no longer guarantee the average AoI performance with fairness, since it indicates $\sum^I_{i=1} m^\circ_{r,i} > M^\circ$. In other words, we cannot obtain the optimal scheduling policy without solving Problem~\eqref{problem:reformulation} as a whole.  
Therefore, we are interested in a low-complexity solution. To this end, we let $M^\circ=Im^\circ_{r,i}$ instead. Then, we follow the same steps as before by solving  $I$ independent convex optimization problems to obtain the scheduling policy $\pi^\circ$, which is a sub-optimal solution for Problem~\eqref{problem:reformulation}. It also has a low complexity of  $\mathcal{O}(I\log I+4I)$. 

This algorithm can be intuitively interpreted as follows: we find the optimal "free space" $m^\circ_{c}$ for the device $i_{\min}$. Then, we insert other devices one by one with $m^\circ_{r,i}$ which occupies $m^\circ_{c}$. If $m^\circ_{r,i}$ is fully occupied during the process, i.e., the cluster is saturated, we sacrifice the AoI performance by extending the "free space" to fit every device. A pseudocode of the algorithm is shown in Alg.~\ref{alg:proposed}.   

With these results, we are able to answer to second key question in Section~\ref{sec:system_model}: The number of devices a cluster can support can be characterized with cluster capacity $C_{\text{cap}}$, with which a low-complexity Algorithm is proposed in Alg.~\ref{alg:proposed} to obtain the scheduling policy.  
\begin{algorithm}[!t]
    \caption{Efficient solver to~\eqref{problem:reformulation}}\label{alg:proposed}
    \begin{algorithmic}[1]
        \State \textbf{Initial:} $z_1,\dots,z_I, \varepsilon_{\max}$
        \State Let $i_{\min}=\arg\min_i\{z_i\}$
        \State Solve Problem~\eqref{problem:single} and get $(m^\circ_{c,i_{\min}}, m^\circ_{r,i_{\min}})$.
        \State Let $M^\circ=m^\circ_{c,i_{\min}}+m^\circ_{r,i_{\min}}$ and $C_{cap}=\lfloor \frac{M^\circ}{m^\circ_{r,i_{\min}}} \rfloor$.

     \If{ $C_{cap}< I$}
        \For {$j=1,\dots ,I$}
            \State Solve Problem~\eqref{problem:single_j} and get $(m'_{c,j},m'_{r,j})$.
            \If{$j\neq i$}
                \State Let $m^\circ_{r,j}=\max\{m'_{r,j},m^\circ_{r,i_{\min}}\}$.
                \State Let $m^\circ_{c,j}=M^\circ-m^\circ_{r,j}$  
            \EndIf
        \EndFor
    \Else
        \State Let $M^\circ=Im^\circ_{r,i}$
            \For {$j=1,\dots ,I$}
            \State Solve Problem~\eqref{problem:single_j} and get $(m'_{c,j},m'_{r,j})$.
            \If{$j\neq i$}
                \State Let $m^\circ_{r,j}=\max\{m'_{r,j},m^\circ_{r,i_{\min}}\}$.
                \State Let $m^\circ_{c,j}=M^\circ-m^\circ_{r,j}$  
            \EndIf
        \EndFor
     \EndIf 
    \end{algorithmic}
    \end{algorithm}

\section{Numerical Simulations}
\label{sec:simulations}
In this section, we provide the numerical results to validate our analytical findings and investigate the system performance in the considered scenarios. 
To demonstrate the advantage of our approaches, we also show the performance of benchmarks under the same setups. 

\subsection{Simulation and Benchmark Setups}
Unless specifically mentioned otherwise, we have the following setups for the simulations:
{We consider the system is operated at the carrier frequency of $f=2.4$ GHz with the bandwidth of $10$ MHz. The transmit power from the server is set to $p_c=30$ dBm and the EH efficiency is $\mu=0.5$. We set the residual loop interference $h_I=-104$ dBm and noise power level $\sigma^2=-174$ dBm. Moreover, the path-loss exponent is $\eta=2.7$. For each update, the packet size is set as $D=128$ bits. The devices are randomly distributed within the range of $d_i\in [0.8, 1.6]$ m to the server with a number of $I=30$.} 

We also provide the performance of the following benchmarks with such setups:
\begin{itemize}
    \item \emph{Exhaustive Search}: It computes for all possible combinations of the scheduling policy $\pi$ and finds the one that minimizes the maximum of AoI $\Delta^*_{\max}$. It guarantees  global optimality within the searching range. 
    \item \emph{IBL Solutions}: It optimizes the scheduling based on the ideal assumption of infinite blocklength (IBL) codes, i.e., the updates are always reliable at Shannon's capacity. In other words, based on this assumption, the scheduling should be chosen so that the update duration $M$ is minimized while fulfilling the conditions that $\log_2(1+\gamma_i)\geq \frac{D}{m_{r,i}}$, $\forall i\in\mathcal{I}$. We show the FBL performance with the corresponding IBL solutions to demonstrate the motivation of considering the FBL codes in networks. 
\end{itemize}

\subsection{Comparison of AoI and Error probability}

\begin{figure}[t!]
    \centering
    \includegraphics[width=\linewidth,trim = 0 0 0 0]{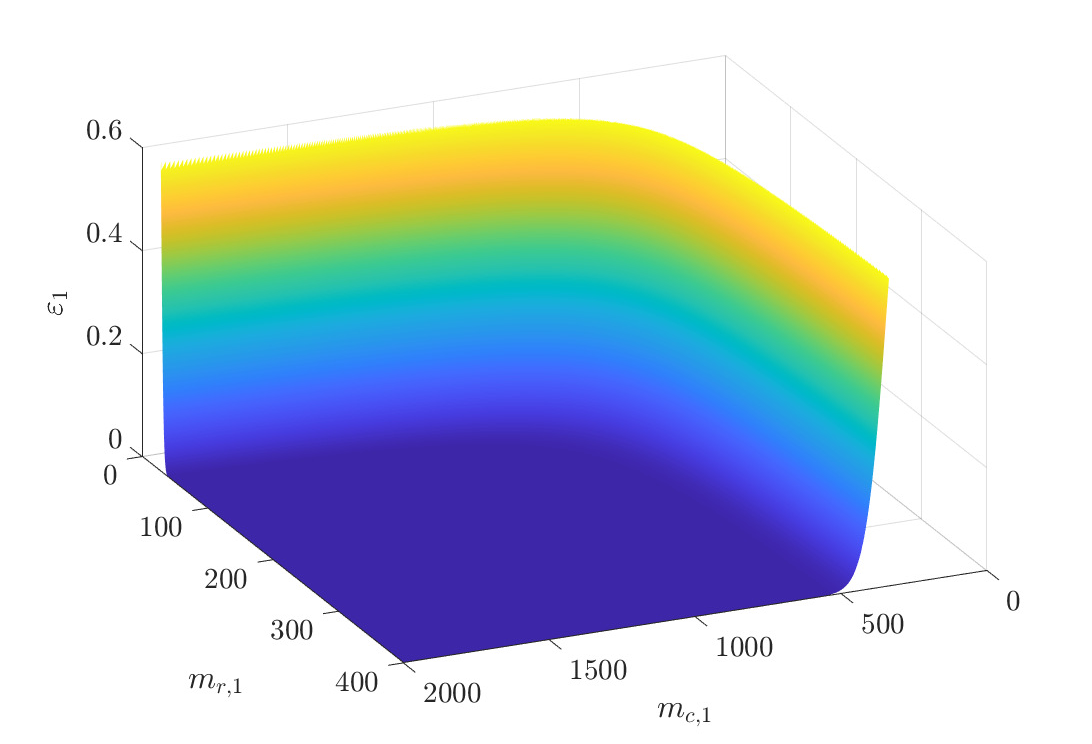}
    \caption{The impact of charging duration $m_{c,1}$ and update duration $m_{r,1}$ on the error probability $\varepsilon_1$, where the device is located at range $d_i=1$ m.}
    \label{fig:err_vs_m_r1_m_c1}
\end{figure}

\begin{figure}[t!]
    \centering
    \includegraphics[width=\linewidth,trim = 0 0 0 0]{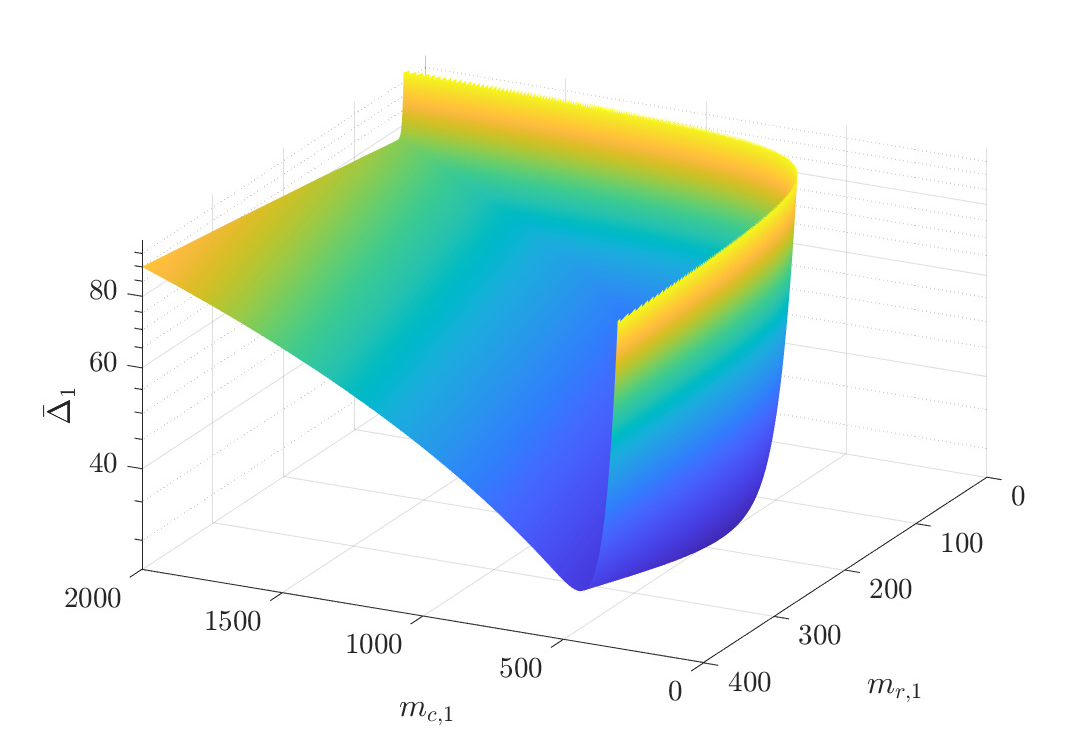}
    \caption{The impact of charging duration $m_{c,1}$ and update duration $m_{r,1}$ on the average AoI $\bar{\Delta}_1$ where the device is located at range $d_i=1$ m.}
    \label{fig:AoI_vs_m_r1_m_c1}
\end{figure}

First, we illustrate the impact of charging duration $m_{c,i}$ and update duration $m_{r,i}$ on the average AoI $\bar{\Delta}_i$ and the error probability $\varepsilon_i$ for the considered device $i$. In particular, we set the first device at range $d_1=1$ m. Then, we plot $\bar{\Delta}_1$ against $m_{c,1}$ and $m_{r,1}$, as well as $\varepsilon_1$ against them, in Fig.~\ref{fig:err_vs_m_r1_m_c1} and Fig.~\ref{fig:AoI_vs_m_r1_m_c1}, respectively. 
Clearly, $\varepsilon_1$ is jointly convex in $m_{r,1}$ and $m_{c,1}$. In fact, $\varepsilon_1$ is essentially a complementary cumulative distribution function (CCDF), i.e., the Q-function, which characterizes the probability that the packet with the given SNR $\gamma_1$ is correctly decoded against the (Gaussian) random noise. Clearly, it is lower-bounded by $0$, and when  $\log_2(1+\gamma_1)\geq \frac{D}{m_{r,1}}$, it becomes convex in  $\omega_1=\sqrt {\frac{m_{r,1}}{V(\gamma_{1})}}  ({\mathcal{C}}(\gamma_1 ) -\frac{D}{m_{r,1}})$, which can be shown concave in $m_{r,1}$ and $m_{c,1}$ in Lemma~\ref{lemma:err_convex}. Note that with the condition of $\gamma_1\geq\gamma_{th}\geq 1$, $\varepsilon_1$ can be improved by both increasing $m_{r,1}$ and $m_{c,1}$. Therefore, if the error probability is the only concern in the system, we should just allocate all available resources, i.e., all symbol lengths, to each update~\cite{Zhu_swipt_2019}. With multiple nodes, it is about addressing the resource balance between nodes, which is already well-investigated, e.g., in~\cite{Pan_URLLC_NOMA_OMA_2020,Zhu_joint_2023}. 

However, in IoT networks, we are more interested in the freshness of the data. When the concern of systems becomes the AoI, it is not always beneficial to have the error probability as low as possible as shown in Fig.~\ref{fig:AoI_vs_m_r1_m_c1}. In particular, the influence of scheduling policy to $\Delta_i$ are two folded: On one hand, longer $m_{c,1}$ provides more harvested energy, i.e, higher SNR with the given $m_{r,1}$ and better $\varepsilon_i$. However, it also means the update round is prolonged, resulting in worse $M$. On the other hand, reducing $m_{r,1}$ implies the increase of energy in each blocklength for the update, i.e., higher SNR with the given $m_{c,1}$. However, it also indicates that the update has less blocklength for the update. Therefore, there exists a tradeoff between $\varepsilon_1$ and $M$, which leads to the quasi-convexity of $\Delta_1$. 
This observation confirms our analytical findings in Corollary~\ref{corollary:problem_convex}. Moreover, the unique characteristic of AoI compared to other conventional metrics also motivates us to investigate its scheduling policy.  

\subsection{Results Validation with Benchmarks}

\begin{figure}[t!]
    \centering
    \includegraphics[width=\linewidth,trim = 0 0 0 0]{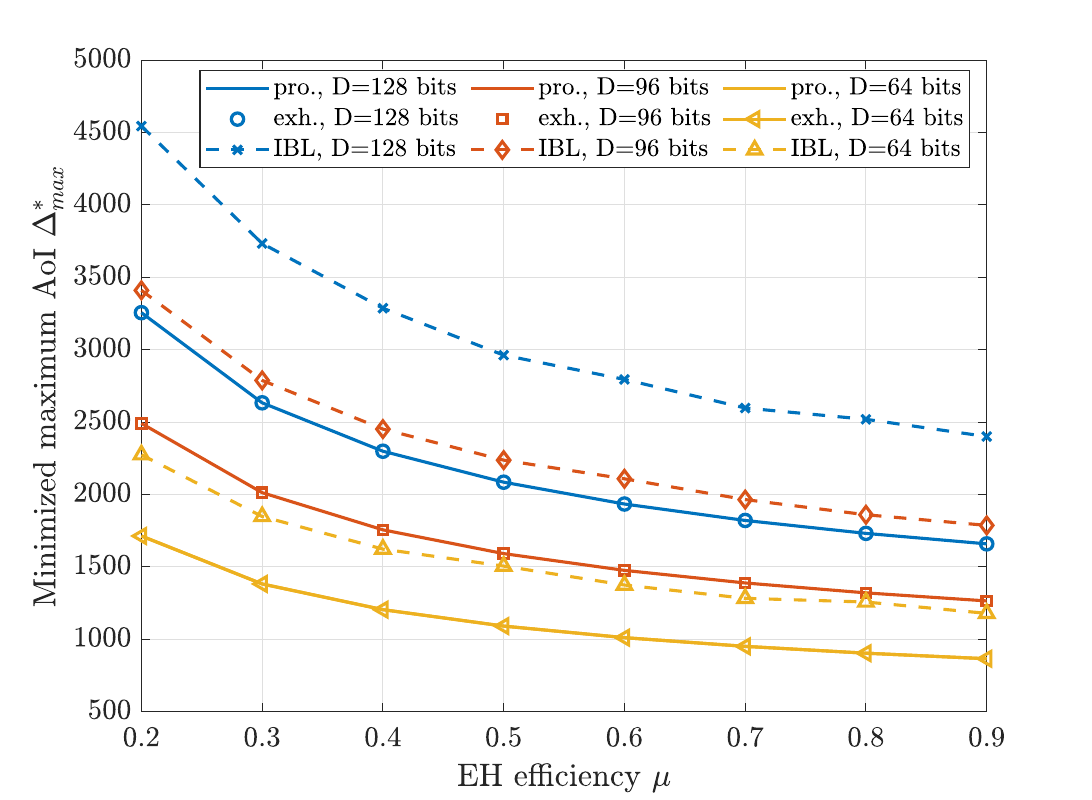}
    \caption{The minimized maximum of AoI among devices $\Delta^*_{\max}$ versus EH efficiency $\mu$ under various setups of $D=\{64, 96, 128\}$ bits. The results obtained by our proposed solution (solid lines) are compared with the results obtained by exhaustive search (markers), as well as the ones obtained with IBL solutions (dash lines).}
    \label{fig:impact_of_pc}
\end{figure}

Next, we show the advantage of our proposed solution by solving Problem~\eqref{problem:reformulation} by comparing the results with two benchmarks. In particular, we plot the minimized maximum of AoI among devices $\Delta^*_{\max}$ versus EH efficiency $\mu$ under various setups of $D=\{64, 96, 128\}$ bits in Fig.~\ref{fig:impact_of_pc}. The results obtained by our proposed solution (indicated as \textbf{pro.}) are depicted with the solid line. Moreover, we also compare these results with two benchmarks, where the ones obtained by exhaustive search (indicated as \textbf{exh.}) are shown with a marker while ones obtained with IBL solutions (indicated as \textbf{IBL}) are plotted with a dash line. 

As expected, $\Delta^*_{\max}$ reduces when we increase $\mu$, since higher $\mu$ indicates more harvested power. Therefore, it requires less charging duration $m_{c,i}$ to achieve the same level of error probability $\varepsilon_i$. Moreover, the improvement becomes flat when $\mu$ is already high due to the fixed received power. It should be pointed out that we may not observe the same behavior when increasing the transmit power $p_c$. This is due to the fact that the server is in the full-duplex mode and suffers from the self-interference, which also scales with $p_c$. 
We can also observe that our proposed solutions are able to achieve the same performance as the ones with the exhaustive search. However, since we solve Problem~\eqref{problem:reformulation} via convex programming, the complexity is much lower. 
On the other hand, the AoI performance with IBL solutions, which ignores the influence of FBL codes, is significantly worse than our proposed solutions. In fact, as discussed in Fig.~\ref{fig:err_vs_m_r1_m_c1} and Fig.~\ref{fig:AoI_vs_m_r1_m_c1}, to obtain the IBL solutions is to choose the scheduling policy so that $\log_2(1+\gamma_i)=\frac{D}{m_{r,i}}$. Under the IBL assumption, it means that the AoI is minimized with no update error, i.e., $\varepsilon_i=0$. However, FBL model in~\eqref{eq:error_tau} indicates that we have $\varepsilon_i=0.5$ if it holds $\log_2(1+\gamma_i)=\frac{D}{m_{r,i}}$. 
Therefore, with FBL codes, if we simply adopt IBL model, the performance will be much worse. This motivates us to revisit the scheduling policy design with the consideration of FBL impact. 

We also plot the minimized maximum of AoI among devices $\Delta^*_{\max}$ versus packet size $D$ under various setups of $\mu=\{0.3, 0.6, 0.9\}$ that are obtained by our proposed solutions, the exhaustive search, and with the IBL solutions in the similar style of Fig.~\ref{fig:impact_of_pc}, respectively. 
We also observe similar trends, i.e., $\Delta^*_{\max}$ increases if $D$ becomes large. Moreover, our proposed solutions can also achieve  global optimality and outperform the results with IBL solutions. However, when $D$ is small, the gap between them becomes insignificant. This is due to the fact that the required power for a low error probability is also small. In fact, $\bar{\Delta}^*_{i}$ is dominated by $M$ if $\varepsilon_i$ approaches to $0$. Although this is also true for increasing $\mu$, its performance is still lower-bounded by the transmit power $p_c$.

\begin{figure}[t!]
    \centering
    \includegraphics[width=\linewidth,trim = 0 0 0 0]{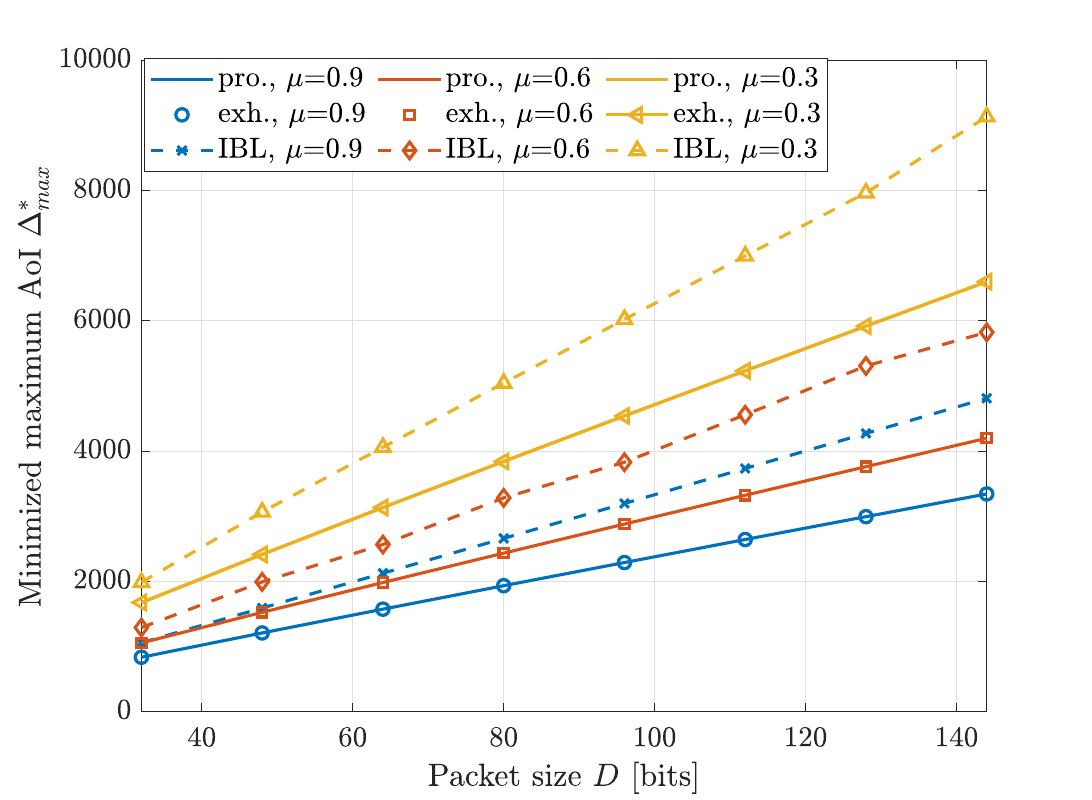}
    \caption{The minimized maximum of AoI among devices $\Delta^*_{\max}$ versus packet size $D$ under various setups of $\mu=\{0.3, 0.6, 0.9\}$. The results obtained by our proposed solution (solid lines) are compared with the results obtained by exhaustive search (markers), as well as the ones obtained with IBL solutions (dash lines).}
    \label{fig:impact_of_D}
\end{figure}

\subsection{Impact of Number of devices}
\begin{figure}[t!]
    \centering
    \includegraphics[width=\linewidth,trim = 0 0 0 0]{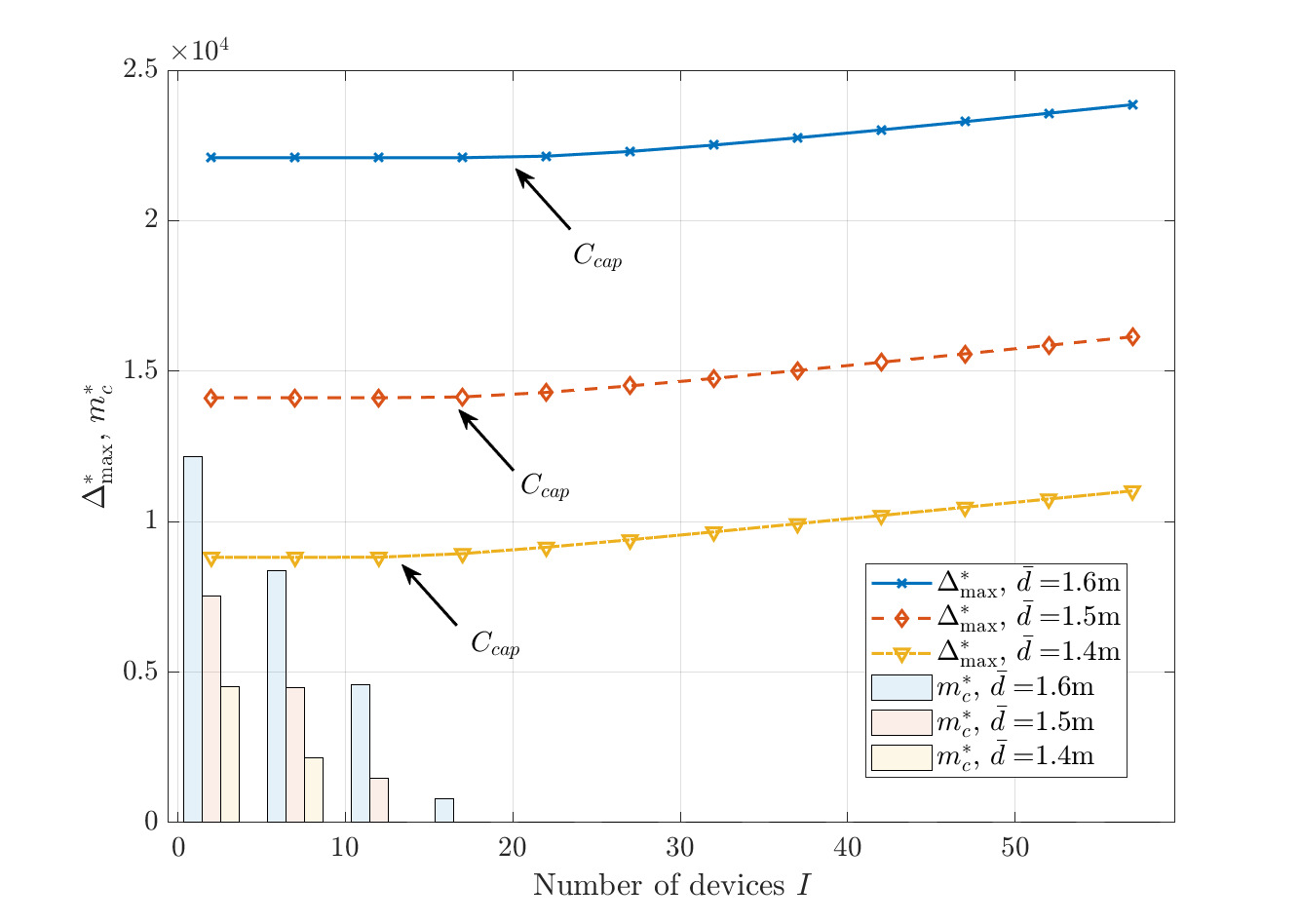}
    \caption{The minimized maximum AoI $\Delta^*_{\max}$ and the optimal common charging duration $m^*_c$ versus the number of devices $I$ under various setups of distance $\bar{d}= \{1.4, 1.5, 1.6\}$ m. }
    \label{fig:impact_of_users}
\end{figure}

In this subsection, we investigate the impact of the number of devices on the AoI performance and the role of cluster capacity in the system design. In particular, we plot the minimized maximum AoI $\Delta^*_{\max}$ (depicted as lines) and the optimal common charging duration $m^*_c$ (depicted as bars) versus the number of devices $I$ under various setups of distance $\bar{d}= \{1.4, 1.5, 1.6\}$ m in Fig~\ref{fig:impact_of_users}. Moreover, for each setup, we indicate the cluster capacity $C_{\text{cap}}$ obtained with~\eqref{eq:capa}. 
For the sake of generality, in this figure, we consider that all devices are homogenous with the unified distance $\bar{d}$. 
When $I\leq C_{\text{cap}}$, $\Delta^*_{\max}$ is not influenced by $I$. This is due to the fact that the "free space", i.e., the common charging duration $m^*_c$ is non-zero. Therefore, the cluster is unsaturated and can support more devices. However, if we keep adding more devices so that $I>C_{\text{cap}}$, $\Delta^*_{\max}$ starts to grow. This is due to the fact that the cluster is now saturated with $m^*_c=0$. 
Clearly, the further distance of the devices is, the worse $\Delta^*_{\max}$ becomes. However, it means that $C_{\text{cap}}$ is also larger since the devices require more energy to carry out a reliable transmission with worse channel gain, i.e., larger $m^*_c$. In other words, the cluster can support more devices without influencing $\Delta^*_{\max}$. 
Therefore, as discussed in Section~\ref{sec:capacity}, $C_{\text{cap}}$ is a metric related to each setup. Its absolute value does not directly indicate the AoI performance.

\begin{figure}[t!]
    \centering
    \includegraphics[width=\linewidth,trim = 0 0 0 0]{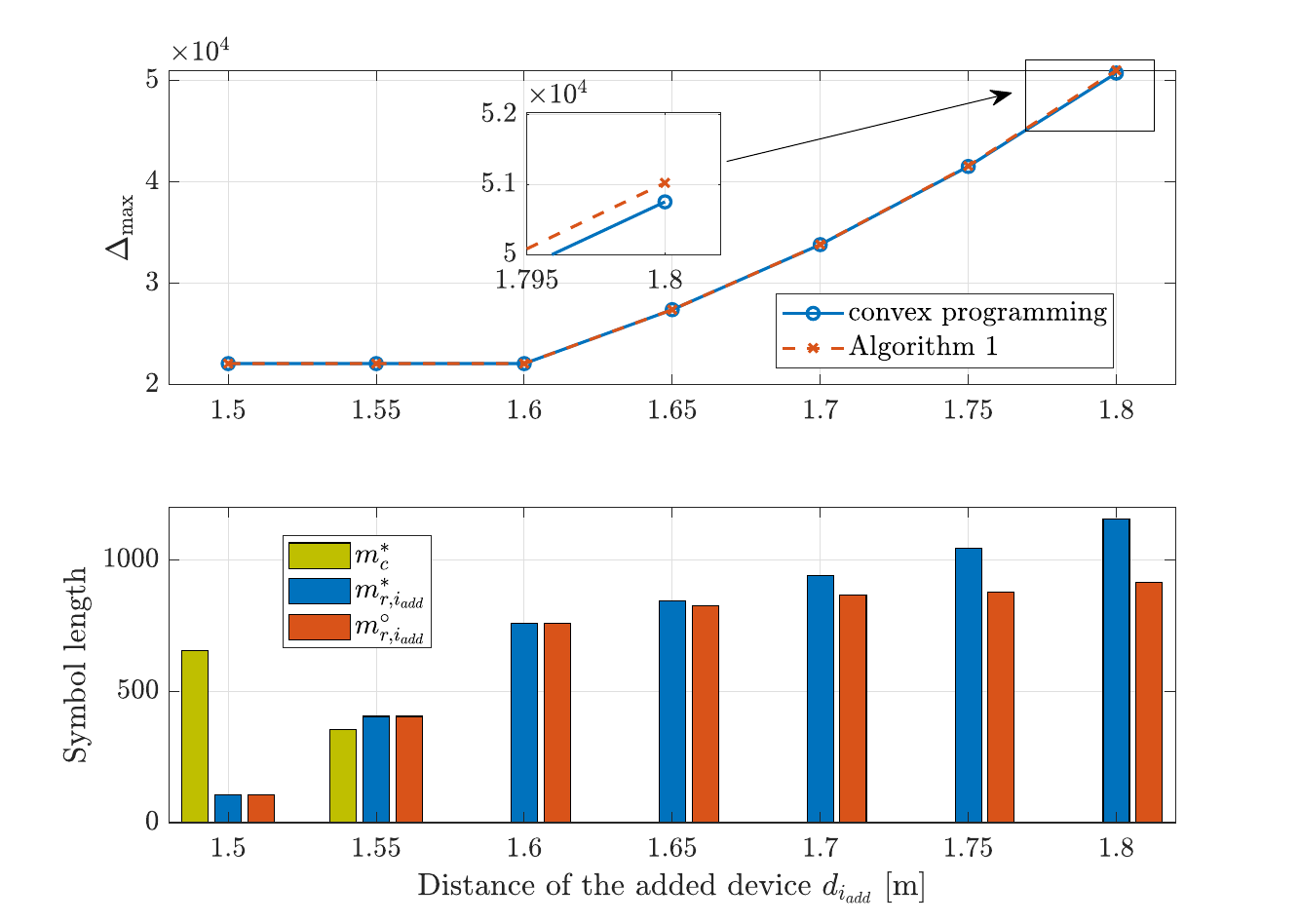}
    \caption{The minimized maximum AoI $\Delta_{\max}$, common charging duration $m^*_c$, and update duration $m^*_{r,i_{\text{add}}}$ versus the distance of added devices $d_{i_{\text{add}}}$ [m] with $16$ existing devices in the distance of $\bar{d}=1.6$ m. Both results from solving convex problem~\eqref{problem:reformulation} and Alg.~\ref{alg:proposed} are shown. }
    \label{fig:impact_of_d_i}
\end{figure}

Since Fig.~\ref{fig:impact_of_users} shows the impact of the number of homogenous devices, it is also interesting to investigate the impact of adding different devices on the AoI performance. Therefore, we set $16$ homogenous devices with distance $\bar{d}=1.6$ m in the cluster, and add an additional devices $i_{\text{add}}$ with distance $d_{i_{\text{add}}}$.
Then, we plot the minimized maximum AoI $\Delta_{\max}$ obtained via convex programming in Problem~\ref{problem:reformulation} versus $d_{i_{\text{add}}}$ in the top sub-figure of Fig.~\ref{fig:impact_of_d_i} while the common charging duration $m^*_c$ and update duration of the added device $m^*_{r,i_{\text{add}}}$ in the bottom sub-figure. Moreover, we also show the AoI obtained with Alg.~\ref{alg:proposed} and the corresponding update duration $m^\circ_{i_{\text{add}}}$ in each sub-figure.    
Similar to the observation in Fig.~\ref{fig:impact_of_users}, $\Delta_{\max}$ remains unchanged when it holds $d_{i_{\text{add}}}\leq \bar{d}$. It implies that the cluster is unsaturated. However, once the device $i_{\text{add}}$ becomes the furthest one in the cluster, $\Delta_{\max}$ increases. This is due to the fact that $m_{i_{\text{add}}}$ occupies more "free space" than the cluster could provide, which is demonstrated in the bottom sub-figure. Therefore, the AoI performance of the system is always lower-bounded by the AoI of the worse device. 
Moreover, we can observe that our proposed Algorithm in Alg.~\ref{alg:proposed} is able to achieve the globally optimal solution, when the cluster is unsaturated. 
However, if the cluster becomes saturated, we lose the global optimality. That being said, the performance gap between our Algorithm and the optimal solution is acceptable when its distance is not far away from other devices, e.g., within $1$ m in our setups. Considering the significantly low complexity, it can still be applied in practical systems even when the cluster is saturated. This observation confirms the advantage of our algorithm.  

\section{Conclusion}
\label{sec:conclusion}

\textcolor{black}{In this paper, we studied the data collection scenario in IoT networks with a particular focus on mURLLC services with WPT-powered devices. We highlighted the importance of real-time data, using the AoI as a metric indicating the timeliness of data. We formulated a fairness-aware AoI minimization problem by optimizing their update scheduling with the consideration of the influence of FBL codes on the AoI. 
To simplify the problem, we establish an equivalent, less complex scheduling policy. Our analytical findings allowed us to efficiently reformulate and solve the problem as a convex one.  
Additionally, we introduced the concept of AoI-oriented cluster capacity, which answers the key question of how many devices can be supported in the network without affecting the AoI. Our numerical results validated our analytical findings and demonstrated the impact of different parameters, which may provide practical insights for the designs of future IoT system with mURLLC services.}

\appendices

\section{Proof of Lemma~\ref{lemma:err_convex}}
\label{app:Lemma_convex}
First, we introduce an auxiliary function:
 \begin{equation}
    \omega_i=\sqrt{\frac{m_{r,i}}{V(\gamma_i)}} (\mathcal{C}(\gamma_i)-\frac{{\color{black}D}}{m_{r,i}})ln2,
 \end{equation}
 with which we have $\varepsilon_i=Q(\omega_i)$. 
 Then, we investigate the convexity of $\varepsilon_i$ with respect to each single variable.  
In particular, the second derivative of $\varepsilon_i$ with respect to $m_{c,i}$ is given by:
\eqsplit{
        \frac{\partial^2 \varepsilon_i}{\partial m_{c,i}^2} &= \underbrace{\frac{\partial^2 \varepsilon_i}{\partial \omega_i^2}}_{\geq 0} \underbrace{\left( \frac{\partial \omega_i}{\partial m_{c,i}} \right)^2}_{\geq 0} + \underbrace{\frac{\partial \varepsilon_i}{\partial \omega_i}}_{\leq 0} \cdot \underbrace{\frac{\partial^2 \omega_i}{\partial m_{c,i}^2}}_{\leq 0}\geq 0.    
} 
The inequality holds since $\frac{\partial^2 \varepsilon_i}{\partial \omega_i^2}={\color{black}\frac{1}{\sqrt{2\pi}}\omega_ie^{-\frac{\omega_i^2}{2}}}$ and $\frac{\partial \varepsilon_i}{\partial \omega_i}={\color{black}-\frac{1}{\sqrt{2\pi}}e^{-\frac{\omega_i^2}{2}}}$. 
Moreover, according to~\cite{Zhu_joint_2023}, we have  $\frac{\partial^2 \omega_i}{\partial m_{c,i}^2}=\frac{\partial^2 \omega_i}{\partial \gamma^2}\left(\frac{\mu_iz^2_iP_c}{\sigma^2+h_Ip_c}\right)^2\leq 0$. Hence, $\varepsilon_i$ is convex in $m_{c,i}$.

Similarly,  the second derivative of $\varepsilon_i$ with respect to $m_{c,i}$ is given by:
\eqsplit{
        \frac{\partial^2 \varepsilon_i}{\partial m_{c,i}^2} &= \underbrace{\frac{\partial^2 \varepsilon_i}{\partial \omega_i^2}}_{\geq 0} \underbrace{\left( \frac{\partial \omega_i}{\partial m_{c,i}} \right)^2}_{\geq 0} + \underbrace{\frac{\partial \varepsilon_i}{\partial \omega_i}}_{\leq 0} \cdot \frac{\partial^2 \omega_i}{\partial m_{c,i}^2}.    
}
Then, we could have $\frac{\partial^2 \varepsilon_i}{\partial m_{c,i}^2}\geq0 $, if it holds $\frac{\partial^2 \omega_i}{\partial m_{c,i}^2}$. In fact, after some manipulations, we have
\eqsplit{
    \frac{\partial^2 \omega_i}{\partial m_{r,i}^2}
    &=b\cdot(a_0+a_1\gamma_i+a_2\gamma^2_i+a_3\gamma^3_i+a_4\gamma^4_i+a_5\gamma^5_i)\\
    &\triangleq b\cdot f(\gamma_i), 
}
where 
\begin{equation}
    \begin{aligned}
    a_0&=12\ln2 r_i-36r_i,\\
    a_1&= 8-5\ln(1+\gamma_i)-126r_i+39\ln2r_i,\\ 
    a_2&=20-162r_i-16\ln(1+\gamma_i)+42ln2r_i,\\
    a_3&=16-90r_i-18\ln(1+\gamma_i)+12ln2r_i,\\ 
    a_4&=4-8\ln(1+\gamma_i)-18r_i-6ln2r_i,\\
    a_5&=-\ln(1+\gamma_i)-3\ln2r_i.
    \end{aligned}
\end{equation}
    Moreover, we have 
\begin{equation}
    b_0=\frac{\ln2m^6_{r,i}}{4bm_{r,1}ln2(b+2m_{r,1})^4(\frac{m_{r,1}^3+2b m_{r,1}^2+b^2m_{r,1}}{b^2+2bm_{r,1}})^{3/2}}\geq 0, 
\end{equation}
where $b = \frac{z_{r,i}z_{c,i}u_ip_cm_{c,i}}{\sigma^2+h_Ip_c}$ and $r_i=\frac{{\color{black}D}}{m_{r,i}}$. 
Note that $f(\gamma_i)$ is a polynomial. We can establish further inequalities to facilitate its expression:
\begin{equation}
    \label{eq:poly_a_5}
    a_5\gamma_i^5 \leq (-\ln(1+\gamma_i)-3\ln2r_i)\gamma_i^2
\end{equation}
\begin{equation}
    \label{eq:poly_a_4}
    a_4\gamma^4_i\leq (4-8\ln(1+\gamma_i)-18r_i)\gamma_i^2-6\ln2r_i\gamma_i
\end{equation} 
\begin{equation}
    \label{eq:poly_a_3}
    a_3\gamma_i^3\leq (16-18\ln(1+\gamma_i)-87r_i+12\ln2r_i)\gamma_i^3-3r_i\gamma_i 
\end{equation}
Combing~\eqref{eq:poly_a_5} -~\eqref{eq:poly_a_3}, we can reduce the order of $f$ with an inequality:
\eqsplit{
    f(\gamma_i)\leq \hat{a}_3\gamma_i^3+\hat{a}_2\gamma_i^2+\hat{a}_1\gamma_i+\hat{a}_0.
}
Therefore, we have $f\leq 0$, if
\begin{equation}
    \begin{aligned}
    \hat{a}_3=16-18ln(1+\gamma)-87r+12lnr\leq 0\\
    \hat{a}_2=24-180r-25ln(1+\gamma)+39ln2r \leq 0\\
    \hat{a}_1=8-129r-5ln(1+\gamma)+33ln2r \leq 0\\
    \hat{a}_0=12ln2r-36r \leq 0.
    \end{aligned}
\end{equation}
As a result, $\varepsilon_i$ is convex in $m_{r,i}$ if $r_i\geq \frac{16-18ln(1+\gamma_i)}{87-12ln2} \geq\frac{16-18ln2}{87-12ln2}=0.0449$

Next, we move on to the joint convexity. The Hessian matrix of $\varepsilon_i$ is given by:
\begin{equation}
    \begin{gathered}
    \mathbf{H} = 
        \begin{bmatrix}
            \frac{\partial^2 \varepsilon_i}{\partial m_{r,i}^2} & \frac{\partial^2 \varepsilon_i}{\partial m_{r,i}\partial m_{c,i}}
        \\  \frac{\partial^2 \varepsilon_i}{\partial m_{c,i}\partial m_{r,i}} &     \frac{\partial^2 \varepsilon_i}{\partial m_{c,i}^2}
        \end{bmatrix}    
    \end{gathered}
\end{equation}
where 
\begin{equation}
    \frac{\partial^2 \varepsilon_i}{\partial m_{r,i}\partial m_{c,i}} = \frac{\partial^2 \varepsilon}{\partial \gamma^2}\frac{\partial \gamma}{\partial m_{c,i}}\frac{\partial \gamma}{\partial m_{r,i}}+\frac{\partial \varepsilon_i}{\partial \gamma}\frac{\partial \gamma}{\partial m_{c,i} \partial m_{r,i}},
\end{equation}
\begin{equation}
        \frac{\partial^2 \varepsilon_i}{\partial m_{c,i}^2} = \frac{\partial^2 \varepsilon_i}{\partial \gamma^2}(\frac{\partial \gamma}{\partial m_{c,i}})^2 + \frac{\partial \varepsilon_i}{\partial \gamma}\frac{\partial^2 \gamma}{\partial m_{c,i}^2} + \frac{\partial^2 \varepsilon_i}{\partial m_{c,i}^2},
\end{equation}
and
\begin{equation}
    \frac{\partial^2 \varepsilon_i}{\partial m_{r,i}^2} = \frac{\partial^2 \varepsilon_i}{\partial \gamma^2}(\frac{\partial \gamma}{\partial m_{r,i}})^2 + \frac{\partial \varepsilon_i}{\partial \gamma}\frac{\partial^2 \gamma}{\partial m_{r,i}^2} + \frac{\partial^2 \varepsilon_i}{\partial m_{r,i}^2}.
\end{equation}
As we showed before, the upper-left element of $\mathbf{H}$ is non-negative. Moreover, its determinate can be written as: 
\eqsplit{
    \det[H] &= \frac{\partial^2 \varepsilon_i}{\partial m_{c,i}^2} \frac{\partial^2 \varepsilon_i}{\partial m_{r,i}^2} - (\frac{\partial^2 \varepsilon_i}{\partial m_{r,i}\partial m_{c,i}})^2 \\
    &= \frac{\partial^2 \varepsilon_i}{\partial \gamma_i^2}\frac{\partial^2 \varepsilon_i}{\partial m_{r,i}^2}(\frac{\partial \gamma_i}{\partial m_{c,i}})^2 - (\frac{\partial \varepsilon_i}{\partial \gamma_i}\frac{\partial \gamma_i}{\partial m_{r,i}\partial m_{c,i}})^2   \\
    &= \frac{P^2}{m_{r,i}^2\sigma_s^4} \bigg(\frac{\partial^2 \varepsilon_i}{\partial \gamma_i^2}\frac{\partial^2 \varepsilon_i}{\partial m_{r,i}^2}-\frac{\partial \varepsilon_i}{\partial \gamma_i}\frac{1}{m_{r,i}^2}\bigg)\\
    &\geq \frac{P^2}{m_{r,i}^2\sigma_s^4} \bigg(\frac{\partial^2 \varepsilon_i}{\partial \omega_i^2}(\frac{\partial \omega_i}{\partial \gamma_i})^2\frac{\partial \varepsilon_i}{\partial \omega_i}\frac{\partial^2 \omega_i}{\partial m_{r,i}^2}-(\frac{\partial \varepsilon_i}{\partial \omega_i}\frac{\partial \omega_i}{\partial \gamma_i})^2\bigg)\\
    &= \frac{P^2}{m_{r,i}^2\sigma_s^4}\bigg(\frac{\partial \omega_i}{\partial \gamma_i}\bigg)^2\frac{\partial \varepsilon_i}{\partial \omega_i}\frac{1}{\sqrt{2\pi}}e^{-\frac{\omega_i^2}{2}}\bigg(\omega_i\frac{\partial^2 \omega_i}{\partial m_{r,i}^2}+\frac{1}{m_{r,i}^2}\bigg)\\
    &=A\bigg(\sqrt{\frac{m_{r,i}}{V}}(C-\frac{{\color{black}D}}{m_{r,i}})\\
    &~~~~ \cdot \big(-\frac{C}{4\sqrt{m_{r,i}^3V}}-\frac{3}{4}\frac{{\color{black}D}}{m_{r,i}\sqrt{m_{r,i}^3V}}\big)ln2+\frac{1}{m_{r,i}^2}\bigg)\\
    &\overset{V\leq1,\varepsilon\leq0.1}{\geq}A\bigg(1.25(-\frac{C}{4m_{r,i}}-\frac{3}{4}\frac{{\color{black}D}}{m_{r,i}^2})ln2+\frac{1}{m_{r,i}^2}\bigg)\\
    &\geq 0,
} 
where $A=\frac{P^2}{m_{r,i}^2\sigma_s^4}\bigg(\frac{\partial \omega_i}{\partial \gamma_i}\bigg)^2\frac{\partial \varepsilon_i}{\partial \omega_i}\frac{1}{\sqrt{2\pi}}e^{-\frac{\omega_i^2}{2}}\geq 0$. 
The last inequality holds if 
\begin{equation}
        \mathcal{C}m_{r,i}+3 {\color{black}D}\geq \frac{4}{\ln(2)}.
    \end{equation}
    Hence, $\varepsilon_i$ is jointly convex in $m_{r,i}$ and $m_{c,i}$ if the condition~\eqref{eq:convex_condition} is fulfilled. 
\bibliographystyle{IEEEtran}
\bibliography{reference_nano}

\end{document}